\documentclass[12pt]{revtex4}
\usepackage{graphicx}
\usepackage{amsmath}
\usepackage{amsfonts}
\usepackage{amssymb}

\begin{document}

\title{Variational Principle of Hydrodynamics and Quantization by Stochastic
Process}
\author{T. Kodama and T. Koide}
\affiliation{Universidade Federal do Rio de Janeiro, Rio de Janeiro, Brazil}

\begin{abstract}
The well-known hydrodynamical representation of the Schr\"{o}dinger equation
is reformulated by extendinging the idea of Nelson-Yasue's stochastic
variational method. The fluid flow is composed by the two stochastic
processes from the past and the future, which are unified naturally by{\ the
principle of maximum entropy}. We show that This formulation is easily
applicable to the quantization of scalar fields.
\end{abstract}

\maketitle

\section{Introduction}

It is almost a century since the present form of quantum mechanics has been
established. It is one of the most beautiful and sound theories from the
mathematical point of view. Because of the recent development of
experimental technology, we are now possible to perform observations related
to fundamental aspects in quantum mechanics \cite{exp}.

So far, there is no experimental observation that throws doubt on its
validity, but we may encounter a situation where the framework of quantum
mechanics will be forced to be modified in future. In fact, the unification
of quantum mechanics and general relativity is still unknown. Therefore, it
is important to deepen understandings of the foundation of quantum mechanics.

Of course such efforts have already been done from very early stage of the
formulation of quantum mechanics, as de Broglie, Bohm and Vigier \cite%
{Holland}. In 1966, Nelson \cite{Nelson} considered that the quantum
fluctuation can be regarded as the Bernstein-type stochastic behavior of a
particle trajectory and and succeeded in deriving the Schr\"{o}diger
equation \cite{Fey}. Later, Yasue \cite{Yasue} \cite{Morat} reformulated
this idea in the form of optimization of an action for the stochastic
variables, known as stochastic variational method (SVM). Several
applications of Yasue's SVM have been found in Ref. \cite{KK,KK2} and
references therein.

In this paper, we propose an alternative representation of the stochastic
quantization within the variational approach. In the present formulation, we
first require the maximum of the entropy associated with stochastic
trajectories in order to accomodate consistently the forward and backward
stochastic processes in the Nelson-Yasue approach. We then show that the
action of the system can be expressed as that of a classical ideal fluid
with the quantum correction as the internal energy of the fluid element \cite%
{jackiw}. The variation of our action can be cast into the form of the
action principle of the well-known quantum mechanics if we change the
variables adequately.

The present paper is organized as follows. In Sec. II, for the sake of
book-keeping, the classical action principle for one particle system and its
generalization to the statistical ensemble are summarized. In Sec.III, we
introduce the two Brownian motions necessary in the variational formulation
and postulate the maximum entropy principle. In Sec.IV, the action of this
combined fluid is established. From this action, we show that it is always
possible to construct the linear representation of the dynamics in term of
the wave function $\psi .$ In Sec.\ V, we discuss the correspondence to the
momentum and Hamiltonian as the usual operator forms and the significance of
eigenstates and eigenvalues through Noether's theorem. In Sec. VI,
application of the present appoach to the field quantization is discussed.
Sec.VII is devoted to the summary and discussion.

\section{Classical single-particle system and Complex representation of its
Action}

As is well-known, the usual classical Lagrangian of a one-particle system is
given by 
\begin{equation}
L_{p}\left( \mathbf{r},\frac{d\mathbf{r}}{dt}\right) =\frac{m}{2}\left( 
\frac{d\mathbf{r}}{dt}\right) ^{2}-V\left( \mathbf{r}\right) ,
\end{equation}
where $\mathbf{r}$ and $V$ are the particle trajectory and the potential,
respectively. For the sake of later convenience, we express Lagrangian in
terms of hydrodynamic variables.

Let us now consider a set of an infinite number of systems whose dynamics
are equivalent each other and the initial and final conditions are specified
as distribution functions of particles and velocities. That is, we are
thinking of a gas of collisonless particles (dust) under the influence of
the potential. In this ensemble, the dynamical variables are given by the
particle distribution function $\rho\left( \mathbf{r},t\right) $ and the
velocity field, $\mathbf{v}\left( \mathbf{r},t\right) $, instead of the
particle trajectory $\mathbf{r}(t)$. This is the Euler-coordinate
representation of one-particle system which is common in the argument of
hydrodynamics. This Lagrangian density is given by 
\begin{equation}
L_{h}=\frac{1}{2}m\rho\mathbf{v}^{2}-\rho V+\kappa\lambda\left( \mathbf{r}%
,t\right) \left[ \frac{\partial\rho}{\partial t}+\nabla \cdot\left( \rho%
\mathbf{v}\right) \right] ,  \label{Lq_1}
\end{equation}
where the last term represents the dynamical constraint associated with the
conservation of the particle number, 
\begin{equation}
\frac{\partial\rho}{\partial t}+\nabla\cdot\left( \rho\mathbf{v}\right) =0,
\label{Continuity}
\end{equation}
and $\lambda\left( \mathbf{r},t\right) $ is a field corresponding to the
Lagrangian multiplier. Here we introduce a constant $\kappa$ to make $%
\lambda $ an adimensional quantity. Thus, $\kappa$ has the dimension of $%
\hbar$.

The variations for $\mathbf{v}$ and $\rho $ lead to the following equation, 
\begin{equation}
\frac{\partial }{\partial t}\mathbf{v}+\left( \mathbf{v}\cdot \nabla \right) 
\mathbf{v}=-\frac{1}{m}\nabla V,  \label{Euler-2}
\end{equation}%
where 
\begin{equation}
\mathbf{v}(\mathbf{x},t)=\frac{\kappa }{m}\nabla \lambda (\mathbf{x},t).
\label{lambda-v}
\end{equation}%
In the one-particle system, $\rho $ is given by Dirac's delta function along
the particle trajectory and then the above equation is reduced to Newton's
equation of motion.

Eq.(\ref{Euler-2}) can be interpreted as the Euler equation for a hypothetic
\textquotedblleft fluid" which consists of non-interacting particles. In
fact, if we replace the potential $V$ by $U$ which contains the contribution
of an internal energy, the potential gradient $\nabla V/m$ is replaced by
the pressure gradient $\nabla P/\rho $ as is the case of the Euler equation.

We can reduce the number of variables of variation by using Eq. (\ref%
{lambda-v}) in Eq. (\ref{Lq_1}) as 
\begin{equation}
I_{h}\left[ \rho,\lambda\right] =\int_{t_{I}}^{t_{F}}dt\int d^{3}\mathbf{r}\
\rho\left\{ -\frac{\kappa^{2}}{2m}\left( \nabla\lambda\right) ^{2}-V-\kappa%
\dot{\lambda}\right\} ,  \label{ReducedAction}
\end{equation}
where $t_{I}(t_{F})$ denotes the initial (final) time. Note that the same
can be derived even from the Hamilton-Jacob theory, where $\lambda$ plays
the role of generator $S$ \cite{Holstein}. Here, the last term on the left
hand side appears as the influence of the constraint condition associated
with the continuity equation of $\rho$.

Now, instead of the two scalar variables, $\rho$ and $\lambda,$ let us
introduce a complex variable 
\begin{equation}
\psi=\sqrt{\rho}e^{i\lambda}.  \label{wavefunction}
\end{equation}
Then the above action is further re-expressed as 
\begin{equation}
I_{h}\left( \rho,\lambda\right) =\int_{t_{i}}^{t_{f}}dt\int d^{3}\mathbf{r}\
\psi^{\ast}\left\{ i\kappa\frac{\partial}{\partial t}-\bar {H}\right\} \psi,
\label{ClassicalQMaction}
\end{equation}
where%
\begin{equation*}
\bar{H}=-\frac{\kappa^{2}}{2m}\nabla^{2}+\bar{V},
\end{equation*}
with the potential 
\begin{equation}
\bar{V}=V-\frac{\kappa^{2}}{2m}\left( \nabla\ln\sqrt{\rho}\right) ^{2}.
\label{ClassicV}
\end{equation}
In short, the classical action (\ref{ReducedAction}) can be cast into a
similar form of the quantum mechanical action \cite{Holland}. Of course,
they are still different because of the $\rho-$dependent term in $\bar{V}$.
However, if the potential $V$ contained an additional term, such as the
internal energy of fluid, and if this term cancels out the second term in
Eq.(\ref{ClassicV}), the corresponding classical action would be the same as
that of quantum mechanics.

In the following, we show that the above scenario can in fact be constructed.

\section{Brownian motion}

So far, we considered that the particle trajectory is deterministic, but let
us now suppose that it is stochastic due to some unknown external factor. In
classical fluids, such a fluctuation occurs as the influence of the nature
of fluid elements which consist of many internal degrees of freedom \cite{KK}%
. On the other hand, the origin of the stochasticity in quantum mechanics is
not known. Here, we will not discuss what is this origin, but show that such
a stochasticity can generate the internal energy.

In order to introduce the stochasticity, we suppose that the particle
trajectory obeys a Brownian motion, which is characterized by 
\begin{equation}
d\mathbf{r}=\mathbf{u}_{F}\left( \mathbf{r},t\right) dt+\vec{\xi}_{F},\ \ \
\ dt>0.  \label{SDE1}
\end{equation}%
where $\vec{\xi}_{F}$ is the Gaussian noise defined by the probability
distribution function,%
\begin{equation}
P\left( \mathbf{\xi }\right) =\frac{1}{\sqrt{2\pi \sigma ^{2}}}e^{-\frac{%
\vec{\xi}^{2}}{2\sigma ^{2}}},  \label{Pgzi}
\end{equation}%
where $\sigma ^{2}=\nu \left\vert dt\right\vert $ with $\nu $ is a constant
parameter which characterizes the intensity of the noise and has a dimension
of $L^{2}/T$. We refer Eq.(\ref{SDE1}) as forward stochastic difference
equation (FSDE).

The particle distribution of this Brownian motion is calculated from FSDE
and given by the following Fokker-Planck equation, 
\begin{equation}
\frac{\partial \rho _{F}}{\partial t}+\nabla \cdot \mathbf{j}_{F}=0,
\label{FP}
\end{equation}%
where 
\begin{align}
\rho _{F}\left( \mathbf{r},t\right) & =\left\langle \delta \left( \mathbf{r}-%
\mathbf{r}\left( t\right) \right) \right\rangle _{F},  \label{density} \\
\mathbf{j}_{F}\left( \mathbf{r},t\right) & =\rho _{F}\left( \mathbf{u}%
_{F}-\nu \nabla \ln \rho _{F}\right) .  \label{jF}
\end{align}%
Here $\left\langle O\left( t\right) \right\rangle _{F}$ represents the
average of $O$ at the instant $t$ over the whole events of the ensemble
satisfying Eq(\ref{SDE1}).

This equation, for a given $\mathbf{u}_{F}$, could be solved forward in time
as an initial value problem. However, we need to fix not only the initial
but also the final particle distributions in the variational approach. The
solution of the above Fokker-Planck equation does not satisfy a given final
distribution in general.

One possible way to control the final distribution, we think of a stochastic
process backward in time. That is, instead of FSDE (\ref{SDE1}), we consider
the backward stochastic differential equation (BSDE), 
\begin{equation}
d\mathbf{r}=\mathbf{u}_{B}\left( \mathbf{r},t\right) dt+\vec{\xi}_{B},\ \ \
\ dt<0.  \label{SDE2}
\end{equation}
where $\vec{\xi}_{B}$ is again the Gaussian noise as before obeying the
probability distribution function, Eq.(\ref{Pgzi}). The corresponding
Fokker-Plank equation is 
\begin{equation}
\frac{\partial\rho_{B}}{\partial t}+\nabla\cdot\mathbf{j}_{B}=0,  \label{FPB}
\end{equation}
where%
\begin{equation}
\mathbf{j}_{B}=\rho_{B}\left( \mathbf{u}_{B}+\nu\nabla\ln\rho_{B}\right) .
\label{jB}
\end{equation}
Note that the sign of the second term is opposite to Eq. (\ref{jF}).

We should construct a set of stochastic processes which satisfies the both
initial and final conditions by using these two Brownian motions. For this,
first let us consider trajectories which obey FSDE with a given initial
condition and pass in the vicinity of $\mathbf{r}$ at a certain time $t$
which satisfys $t_{I}<t<t_{F}$. The number of the trajectories should be
proportional to $\rho _{F}\left( \mathbf{r},t\right) $. Similarly, the
number of the trajectories which pass the same domain starting from the
final distribution should be proportional to $\rho _{B}\left( \mathbf{r}%
,t\right) $. Therefore, the number $N$ of the ways to construct a trajectory
which combines these two should be proportional to the product of the
densities $N\propto \rho _{F}\left( \mathbf{r},t\right) \rho _{B}\left( 
\mathbf{r},t\right) $.

Now we require that \textit{the law of Nature describes the situation where
this combined number is maximal for any instant} $t$. In other words, the
entropy associated with this combination of trajectories should be maximum.
We then define the entropy 
\begin{equation*}
S[\rho _{F},\rho _{B}]\equiv \int d^{3}\mathbf{r}N\ln N.
\end{equation*}%
From the variation of this entropy $\delta S=0$ with the following
constraint conditions, 
\begin{equation}
\int d^{3}\mathbf{r}\rho _{F}\left( \mathbf{r},t\right) =\int d^{3}\mathbf{r}%
\ \rho _{B}\left( \mathbf{r},t\right) =1,
\end{equation}%
we obtain 
\begin{equation}
\rho _{F}\left( \mathbf{r},t\right) =\rho _{B}\left( \mathbf{r},t\right) .
\label{rhoF=rhoB}
\end{equation}%
Therefore the density of trajectories $\rho $ which satify the two boundary
conditions is given by%
\begin{equation}
\rho \left( \mathbf{r},t\right) =\rho _{F}\left( \mathbf{r},t\right) =\rho
_{B}\left( \mathbf{r},t\right) .
\end{equation}

Once we establish Eq.(\ref{rhoF=rhoB}), we get from Eqs.(\ref{FP},\ref{FPB}) 
\begin{equation}
\rho \left( \mathbf{u}_{F}-\nu \nabla \ln \rho \right) -\rho \left( \mathbf{u%
}_{B}+\nu \nabla \ln \rho \right) =\nabla \chi +\nabla \times \mathbf{A,}
\end{equation}%
where $\mathbf{A}$ is an arbitrary time-dependent vector field \cite{KK} and 
$\chi $ is a scalar function which satisfies $\nabla ^{2}\chi =0.$ In the
presence of $\mathbf{A},$ the velocity field should contain a vortex which
is related to a singularity in space. Similarly, since $\lim_{|\mathbf{r}%
|\rightarrow \infty }\nabla \chi \rightarrow 0$, $\chi =const.\ $, if $\chi $
has no singularity. Here, for simplicity, we assume $\nabla \chi =\mathbf{A}%
=0$. Then we obtain 
\begin{equation}
\mathbf{u}_{F}-\mathbf{u}_{B}=2\nu \nabla \ln \rho .  \label{Consistency}
\end{equation}%
This is nothing but the same relation called the consistency condition in
SVM \cite{KK, Yasue}.

Because of this condition, the two Fokker-Planck equations are reduced to
the simple continuity equation, 
\begin{equation}
\partial _{t}\rho +\nabla (\rho \mathbf{u}_{T})=0,  \label{FP-true}
\end{equation}%
where 
\begin{equation}
\mathbf{u}_{T}=\frac{\mathbf{u}_{F}+\mathbf{u}_{B}}{2}.
\end{equation}

\section{Action principle and quantum mechanics}

In the previous section, we introduce the stochastic process which is
characterized by the two Brownian motions. As is in the hydrodynamic
variational approach, the properties of these Brownian motions can be
incorporated as the constraint condition in the variational approach. That
is, once we know the kinetic term $T$ and potential term $U$, the action
which we should optimize is given by 
\begin{equation*}
I_{q}=\int dt\int d^{3}\mathbf{r}\ \rho \left\{ T-U-\kappa \dot{\lambda}%
-\kappa \nabla \lambda \cdot \mathbf{u}_{T}\right\} .
\end{equation*}%
Here the Fokker-Planck equations (\ref{FP-true}) is taken into account as
the constraint conditions with a Lagrange multipliers, $\lambda $.

From Eq.(\ref{FP-true}), we can identify the velocity $\mathbf{u}_{T}$ with
that of the translational motion of the fluid. Therefore, the translational
kinetic energy $T$ should be 
\begin{equation}
T=\frac{m}{2}\mathbf{u}_{T}^{2}.
\end{equation}

On the other hand, the relative velocity, 
\begin{equation}
\mathbf{u}_{r}=\mathbf{u}_{F}-\mathbf{u}_{B},  \label{U-rel}
\end{equation}
is irrelevant to the translational motion of the fluid element but should be
associated with its internal energy. Such a situation occurs in the case of
the kinetic derivation of hydrodynamics, where the momentum of microscopic
constituent particles are separated into two parts; one is to the fluid
velocity and the other to the internal energy. Following this idea, the
potential term is expressed as 
\begin{equation*}
U=\frac{m_{eff}}{2}\mathbf{u}_{r}^{2}+V=8\alpha^{2}\nu^{2}m\left( \nabla \ln%
\sqrt{\rho}\right) ^{2}+V,
\end{equation*}
where $m_{eff}=\alpha^{2}m$ is an effective mass, and $\alpha$ is still
undetermined constant. As before, $V$ is the external potential.

By using these results, we arrive at the following expression, 
\begin{equation}
I_{q}\left[ \rho ,\lambda \right] =\int dt\int d^{3}\vec{r}\ \rho \left\{ -%
\frac{1}{2m}\left( \kappa \nabla \lambda \right) ^{2}-U-\kappa \dot{\lambda}%
\right\} .  \label{FinalAction}
\end{equation}%
Since $\kappa $ is an arbitrary constant, we can always choose it so as to
satisfy 
\begin{equation}
\kappa =4\alpha \nu m.  \label{alfa}
\end{equation}%
Therefore, as was done in Sec. II, this can be re-expressed in terms of the
usual wave function as 
\begin{equation}
I_{q}\left[ \rho ,\lambda \right] =\int dt\int d^{3}\vec{r}\ \ \psi ^{\ast
}\left( \mathbf{r},t\right) \left[ i\kappa \partial _{t}-\bar{H}\right] \psi
\left( \mathbf{r},t\right) ,\   \label{ActionRhoLambda}
\end{equation}%
where the wave function is defined by Eq. (\ref{wavefunction}) and 
\begin{equation}
\bar{H}=-\frac{\kappa ^{2}}{2m}\nabla ^{2}+V\left( \mathbf{r}\right) .
\label{H-bar}
\end{equation}

Then the variation for $\rho$ and $\lambda$ of the action (\ref%
{ActionRhoLambda}) leads to 
\begin{equation}
\left[ i\kappa\partial_{t}-\left( -\frac{\kappa^{2}}{2m}\nabla^{2}+V\left( 
\mathbf{r}\right) \right) \right] \psi\left( \mathbf{r},t\right) =0,
\label{LinearRep}
\end{equation}
If the parameter $\kappa$ is identified with $\hbar$, this is nothing but
the Schr\"{o}dinger equation.

It should be noted that our result of the variation becomes non-linear if
the parameter does not satisfy Eq. (\ref{alfa}) \cite{Modular}. However,
because of the ambiguity for the definition of the phase $\lambda $, we can
always find a parameter set satisfying Eq. (\ref{alfa}) and thus a
non-linear dynamics can be cast into the form of the linear equation (see
also \cite{Caticha}). In other words, it seems that the wave function is a
convenient representation but is not necessarily the fundamental element to
construct quantum mechanics. This will be discusses in Sec. VII. In the
following, we will refer this linear representation as $\psi $%
-representation.

\section{Definitions of physical operators and eigenvalues}

In the canonical quantization, the momentum operator is defined so as to
maintain the correspondence principle between the Poisson brakets and
commutators. In the present formulation, it is defined through the Noether
theorem.

The momentum is a conserved quantity associated with the invariance for the
spatial translation. Suppose that the action Eq.(\ref{ActionRhoLambda}) is
invariant under the spatial translation. Then, the corresponding conserved
Noether charge is 
\begin{equation}
\mathbf{P}=\int d^{3}\mathbf{x}\rho\nabla\lambda.  \label{Momentum}
\end{equation}
That is, the event average of $\nabla\lambda$ is a conserved quantity.
Similarly, the invariance for the time translation leads to the Noether
conserved charge as 
\begin{equation}
H=\int d^{3}\mathbf{x}\rho\left[ \frac{m}{2}(\mathbf{u}_{m}^{2}+\alpha^{2}%
\nu^{2}(\nabla\ln\rho)^{2})+V\right] .  \label{energy}
\end{equation}
In $\psi$-representation, these conserved charges can be expressed as 
\begin{align}
\mathbf{P} & =\int d^{3}\mathbf{\vec{r}\ }\psi^{\ast}\left( \mathbf{r}%
,t\right) \frac{\hbar}{i}\nabla\psi^{\ast}\left( \mathbf{r},t\right)
\equiv\langle \hat{\mathbf{P}}\rangle,  \label{Op-Momentum} \\
H & =\int d^{3}\mathbf{\vec{r}\ }\psi^{\ast}\left( \mathbf{r},t\right)
\left( -\frac{\hbar^{2}}{2m}\nabla^{2}+V\right) \mathbf{\ }\psi\left( 
\mathbf{r},t\right) \equiv\langle\bar{H}\rangle.  \label{Op-Energy}
\end{align}
These are usual operator representations of physical quantities in quantum
mechanics. For other observables related to the generator of some kind of
transformation, for example, angular momentum can also be defined in a
similar way.

The eigenvalues and corresponding eigenstates can be introduced following
the classical arguments of von Neumann \cite{Neumann}. Let us consider an
wave function $\psi _{P}$ representing the corresponding expectation value
by $\langle ~~\rangle _{P}$. Suppose that this state satisfies 
\begin{equation}
\langle (\hat{\mathbf{P}}-\langle \hat{\mathbf{P}}\rangle _{P})^{2}\rangle
_{P}=0.
\end{equation}%
Then it is clear that $\psi _{P}$ is the solution of the following equation, 
\begin{equation}
\left( \hat{\mathbf{P}}-\mathbf{p}_{0}\right) \psi _{P}\left( \mathbf{r}%
,t\right) =0.
\end{equation}%
Here $\mathbf{p}_{0}$ is an eigenvalue. By the same argument, we can
construct any eigenvalues and eigenstates of physical observables defined by
the Noether theorem.

\section{Application to Field Quantization}

The stochastic quantization procedure in terms of variational principle in
the previous sections can be extended in a straightforward way for any
system described by a vector variable, say $\vec{\phi}$ in stead of $\mathbf{%
r}$. As an example, we sketch in below how the present formulation can be
applied to a scalar field theory.

To apply our formulation for a scalar field system, we introduce the spatial
lattice representation. In this representation, we can assign the field
configuration $\phi \left( x\right) $ to a vector,%
\begin{equation}
\phi \left( x\right) \rightarrow \vec{\phi},
\end{equation}%
in such a way that the scalar product of two functions%
\begin{equation*}
(f,h)=\int d^{3}xf\left( x\right) h\left( x\right) \rightarrow \Delta ^{3}%
\mathbf{x}\ \ \vec{f}\cdot \vec{h}
\end{equation*}%
with $\Delta ^{3}\mathbf{x}$ being the lattice volume.

In this representation, the classical action for the scalar field $\phi
\left( x\right) \ $can be written as 
\begin{equation}
I_{c.f.}=\int_{t_{i}}^{t_{f}}dt\ \left\{ \frac{1}{2c^{2}}\Delta ^{3}\mathbf{x%
}\left( \frac{d\vec{\phi}}{dt}\right) ^{2}+\frac{1}{2}\Delta ^{3}\mathbf{x\ }%
\vec{\phi}\cdot \mathbf{\Delta }_{\mathbf{x}}\vec{\phi}-V\left( \vec{\phi}%
\right) \right\} ,
\end{equation}%
where $V$ is the potential containing the mass term and $\mathbf{\Delta }_{%
\mathbf{x}}$ is the matrix corresponding to the discretized Laplacian
operator \cite{KK2}. By denoting a formal correspondence%
\begin{align}
\vec{r}& \rightarrow \vec{\phi}, \\
d^{3}\vec{r}& \rightarrow d^{N^{3}}\vec{\phi}\equiv \lbrack D\phi ],
\end{align}%
we can repeat the analogous procedure in the previous sections and we have
finally%
\begin{align}
\lefteqn{I_{q.f.}}  \notag \\
& \hspace*{-1cm}=\int_{t_{i}}^{t_{f}}dt\int [D\phi ]\rho \left[ \phi \right]
\left\{ \ \int d^{3}\mathbf{x}\left( \frac{\left( \kappa c\right) ^{2}}{2}%
\left\{ \left( \frac{\delta \lambda }{\delta \phi \left( x\right) }\right)
^{2}-\left( \ \frac{\delta \ln \sqrt{\rho }}{\delta \phi \left( x\right) }%
\right) ^{2}\right\} +\frac{1}{2}\phi \left( x\right) \nabla ^{2}\phi \left(
x\right) \right) -V\left[ \phi \right] -\kappa \partial _{t}\lambda \right\}
.  \notag \\
&  \label{FieldFuid}
\end{align}%
In the above, we used the notation in the continuum limit as%
\begin{equation}
\frac{1}{\Delta ^{3}\mathbf{x}}\nabla _{\vec{\phi}}\rightarrow \frac{\delta 
}{\delta \phi \left( x\right) },
\end{equation}%
and $\rho \left[ \phi \right] $ represents the functional distribution of
field configuration $\phi $.

As before, in terms of wave functional,%
\begin{equation}
\Psi \left[ \phi \right] \equiv \sqrt{\rho }e^{i\lambda },
\end{equation}%
the above action is rewritten as 
\begin{equation}
I_{q.f.}=\int_{t_{i}}^{t_{f}}dt\int [D\phi ]\ \Psi ^{\ast }\left[ \phi %
\right] \left[ i\kappa c\partial _{ct}-\widehat{\mathcal{H}}\right] \Psi %
\left[ \phi \right] ,  \label{ActionScalarField}
\end{equation}%
where%
\begin{equation}
\widehat{\mathcal{H}}=\int d^{3}\mathbf{x}\left\{ -\frac{\kappa ^{2}c^{2}}{2}%
\left( \frac{\delta }{\delta \phi \left( x\right) }\right) ^{2}-\frac{1}{2}%
\left( \nabla \phi \left( x\right) \right) ^{2}\right\} +V.
\label{FieldHamilton}
\end{equation}%
Taking variations with respect to the two functionals $\rho \left[ \phi %
\right] $ and $\lambda \left[ \phi \right] ,$ we arrive at the functional
Schr\"{o}dinger equation%
\begin{equation*}
i\ \kappa c\ \partial _{ct}\Psi \left[ \phi \right] =\widehat{\mathcal{H}}%
\Psi \left[ \phi \right] .
\end{equation*}%
Note that, for the scalar field system, the corresponding condition to Eq.(%
\ref{alfa}) determines the relation between the noise intensity $\nu $ and
the universal constant, $\kappa c.$

\section{Summary and Discussion}

In this work, we formulate the quantization of one-particle classical system
in terms of the variational approach. There are three requirements in the
present derivation; 1) quantum fluctuation is expressed as stochastic noises
of the Brownian motion, 2) the two stochastic flows from the past and the
future are characterized by the maximum of the entropy associated with the
number of combinatory processes connecting the fixed distributions, and 3)
the action of the system is given by the same form of a classical ideal
fluid, but contains the contribution from the internal energy. This fluid
action can be cast into the action principle for the Schr\"{o}dinger
equation. In addition, the variational form permits us to define operators
associated with physical observables in the $\psi -$representations via
Noether's theorem. The usual definition of eigenvalues and corresponding
eigenstates of the physical observables are consistently defined within this
scheme. In short, the well-known hydrodynamic representation of the Schr\"{o}%
dinger equation is reformulated within the framework of the stochastic
variational method of the Nelson-Yasue approach.

The method developed here is easily applicable to the quantization of scalar
fields. In this case, we consider that the velocity in the functional space
is irrotational. As was pointed out by Takabayashi and Wallstrom, the usual
hydrodynamic representation of the Schr\"{o}dinger equation cannot treat the
cases where the phase of wave functions becomes multi-valued, such as the
eigenstates of the angular momentum unless introducing an additional
condition which requires that the vortex number is quantized \cite%
{Takabayashi} \cite{wall}. However, in the field quantization, the vorticity
in question refers to the flow in the functional space and nothing to do
with multivaluedness of the phase of the wavefunction itself. For the case
of scalar field the hydrodynamic equations in the functional space can be
derived from the variation of Eq.(\ref{FieldFuid}) and the flow in the
functional space can be taken always irrotational. In particular for $V=0,$
we obtain the correct energy eigenvalues as the stationary states without
resort to the $\psi -$ representation in functional space \cite{KK2}.

From the present study we may give rein our imagination for the possible
origin of our quantum noise as the fluctuation of the space-time itself. In
fact, the intensity of the noise for the field quantization is given
uniquely by the universal constant $\kappa c=\hbar c$ as seen from Eqs.(\ref%
{ActionScalarField},\ref{FieldHamilton}). This suggest that the field
variables and the space-time structure were born simultaneously in such a
way that quantum mechanics and relativity have the same origin and are not
to be defined separately. If this is the case, we would need to reconsider
the meaning of quantization of gravity.

We have assumed that the stochasticity is characterized by the Gaussian
white noise. This reminds us the well-known the central limit theorem and
suggests that there may exist another hypermicro-stochastic process which is
reduced to the Gaussian white noise only after taking the central limit
theorem. This aspect seems to be consistent with the maximum entropy
postulate which we have adopted, since the average of large micro-stochastic
process should lead to certain statistical equilibrium.

\bigskip

The authors acknowledge the fruitful discussions with N. Suzuki, K. Tsushima
and P. V\'{a}n. This work has been supported by CNPq, FAPERJ, CAPES and
PRONEX.

\end{document}